\documentclass[preprint,showpacs,amsmath,amssymb,english]{revtex4}

\usepackage{graphicx}
\usepackage{dcolumn}
\usepackage{babel}
\usepackage{color}

\begin{document}

\title{Step by step capping and strain state of GaN/AlN quantum dots studied
by grazing incidence diffraction anomalous fine structure}

\author{J. Coraux$^{1,2,*}$, M.G. Proietti$^{3}$, V. Favre-Nicolin$^{1,2}$, H. Renevier$^{1,2}$,
B. Daudin$^{1}$}
\affiliation{$^{1}$Commissariat à l'Energie Atomique, D$\acute{e}$partement de Recherche Fondamentale sur la Mati$\grave{e}$re Condens$\acute{e}$e, SP2M/NRS, 17 rue
des martyrs, 38054 Grenoble Cedex 9, France.\\
$^{2}$Universit$\acute{e}$ Joseph Fourier, BP 53, 38041, Grenoble Cedex 9, France.\\
$^{3}$Departamento de F$\acute{i}$sica de la Materia Condensada, Instituto de Ciencia de Materiales de Arag$\acute{o}$n, CSIC-Universidad de Zaragoza - c. Pedro Cerbuna 12, 50009 Zaragoza, Spain.\\
$^{*}$Johann.Coraux@cea.fr}

\date{\today}

\begin{abstract}
The investigation of small size embedded nanostructures,
by a combination of complementary anomalous diffraction techniques,
is reported. GaN Quantum Dots (QDs), grown by molecular beam epitaxy
in a modified Stranski-Krastanow mode, are studied in terms of strain
and local environment, as a function of the AlN cap layer thickness,
by means of grazing incidence anomalous diffraction. That is, the
X-ray photons energy is tuned across the Ga absorption K-edge which
makes diffraction chemically selective. Measurement of \textit{hkl}-scans,
close to the AlN $\left(30\bar{3}0\right)$ Bragg reflection, at several
energies across the Ga K-edge, allows the extraction of the Ga partial
structure factor, from which the in-plane strain of GaN QDs is deduced.
>From the fixed-\textbf{\textit{Q}} energy-dependent
diffracted intensity spectra, measured for diffraction-selected iso-strain
regions corresponding to the average in-plane strain state of the
QDs, quantitative information regarding composition and the out-of-plane
strain has been obtained. We recover the in-plane and out-of-plane strains
in the dots. The comparison to the biaxial elastic strain in a pseudomorphic
layer indicates a tendency to an over-strained regime.
\end{abstract}


\maketitle

\section{Introduction}

Most of the interest in the improvement of the growth
techniques for III-V semiconductor nanostructures originates from the fact that
the quantum confinement of carriers leads to unique optoelectronic
performances. The confinement in 1 dimension, that is the growth of
quantum wells (QWs), has been brought under control for a variety
of systems, leading to QWs based optoelectronic devices, such as nitride
QWs laser diodes \cite{Ponce97}. However, due to the large densities
of defects in the III-nitride materials \cite{Wu98}, the emission
efficiency of such devices is strongly altered by the increasing of temperature. As
an alternate way to overcome that difficulty, carriers may be confined
in regions free of any defect \cite{Arakawa82}, such as self-organized quantum
dots (QDs). For reasonable optoelectronic efficiency,
a simultaneous control over the size, size distribution, nucleation
sites, density and structure of the QDs is required. In the InAs/GaAs
system, room temperature ultraviolet lasers \cite{Ledentsov02} could
be achieved once these requirements were satisfied. This article focuses on
the GaN/AlN system, for which the overall
control of the QDs characteristics still remains a challenge. In the last few
years, Molecular Beam Epitaxy (MBE) has recently
been of particular interest with the improvement of QDs density control, using
the Stranski-Krastanow (SK) growth mode \cite{Daudin97,Adelmann04,Brown04}
and derivatives \cite{Gogneau03,Adelmann02,Damilano99}. Size homogeneization of
the QDs was achieved using vertical correlation through strain
fields \cite{Tersoff96,Chamard03,Coraux05a}.

Alternatively, new efforts are made to understand
the effects of the QDs capping by AlN, which strongly modifies the
strain state in the QDs \cite{Daudin05,Coraux05a}, and therefore plays a
decisive role in the modification of the optical properties. Moreover,
this topic raises a fundamental interest regarding the physics of
strain accomodation between a QD and its capping.

Many complementary methods have been applied to
quantitative strain characterization in nanostructures \cite{Stangl04}.
This is true for Grazing Incidence X-Ray Diffraction (GIXRD), which can be
made chemically sensitive when carried out as a function of the energy
across the absorption edge of an element. This technique is known
as anomalous diffraction \cite{Hodeau01} and is suitable (a) to localize some element
in reciprocal space \cite{Hendrickson91}, (b) to identify the local
environment of an atom \cite{Stragier92}, and (c) to determine
the composition of an iso-strain region selected by diffraction \cite{Letoublon04}.

In this article we present a comprehensive strain
analysis of the capping of GaN QDs by AlN. More precisely, we study
the in-plane and out-of-plane strain state in the QDs as a function
of the AlN capping thickness, by means of grazing incidence anomalous
x-ray diffraction at the Ga K-edge (10.367 keV), around the $\left(30\bar{3}0\right)$
reflection.

The MBE preparation of the set of samples to be
analyzed is presented in section \ref{Samples}. The strain characterisation technique, by Grazing Incidence Diffraction
Anomalous Fine Structure (GIDAFS), is detailled in section \ref{GIDAFS}. The strain and
composition informations obtained using different aspects of GIDAFS,
namely Multi-wavelength Anomalous Diffraction (MAD), Extended Diffraction
Anomalous Fine Structure (EDAFS) oscillations and diffraction anomalous
lineshape analysis are given in section
\ref{sec:Multiwavelength-Anomalous-Diffraction},
\ref{sec:Extended-DAFS-analysis}, and \ref{Edge analysis}.

\section{Samples}

\label{Samples}

\begin{figure}
\begin{center}\includegraphics[width=8cm,keepaspectratio]{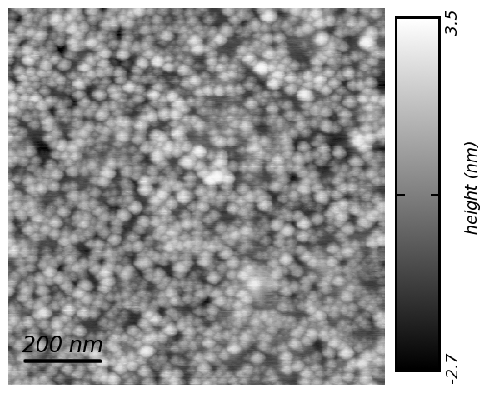}\end{center}
\caption{\label{fig:AFM_2D}}
\end{figure}

The samples were grown in a MECA 2000 MBE chamber,
equiped with standard effusion cells providing the Ga and Al fluxes,
and a radiofrequency plasma cell providing the active nitrogen flux.
The substrate were $2\:\mu m$ thick AlN$\left(0001\right)$ layers
deposited by Metal Organic Chemical Vapor Deposition on sapphire \cite{Shibata03}.
The substrate temperature was fixed at $740^{\circ}C$. Prior to the
growth of the QDs, 10 nm thick AlN buffers were grown. The QDs growth
was achieved in the modified SK growth mode \cite{Gogneau03,Adelmann02},
by depositing 6 GaN monolayers (MLs) under Ga-rich conditions. This
resulted in the formation of a Ga bilayer at the surface inhibiting
the 2D/3D transition even above the usual $\sim$ 2 MLs
critical thickness for the 2D/3D transition in the SK growth mode
\cite{Daudin97}. The thermal evaporation under vacuum of the Ga bilayers
led to the transition of the 2D GaN layer into 3D QDs \cite{Gogneau03,Adelmann02}
connected by a $\sim$ 2 MLs thick Wetting Layer (WL). A
set of 5 samples was grown, with increasing AlN capping : 0,
2, 5, 10, and 20 MLs. Figure \ref{fig:AFM_2D}
shows a $1\:\mu m^{2}$ Atomic Force Microscopy (AFM) image of free standing dots
sample. The height of the QDs was evaluated to $3.0\pm0.5\: nm$, their
diameter to $15\pm1\: nm$, leading to an aspect ratio of about 0.2.
The QDs density was found as high as $1.3\times10^{11}\: cm^{-2}$,
that is QDs are almost adjacent.

\section{GIDAFS measurements}

\label{GIDAFS}

\begin{figure}
\begin{center}\includegraphics[width=12cm,keepaspectratio]{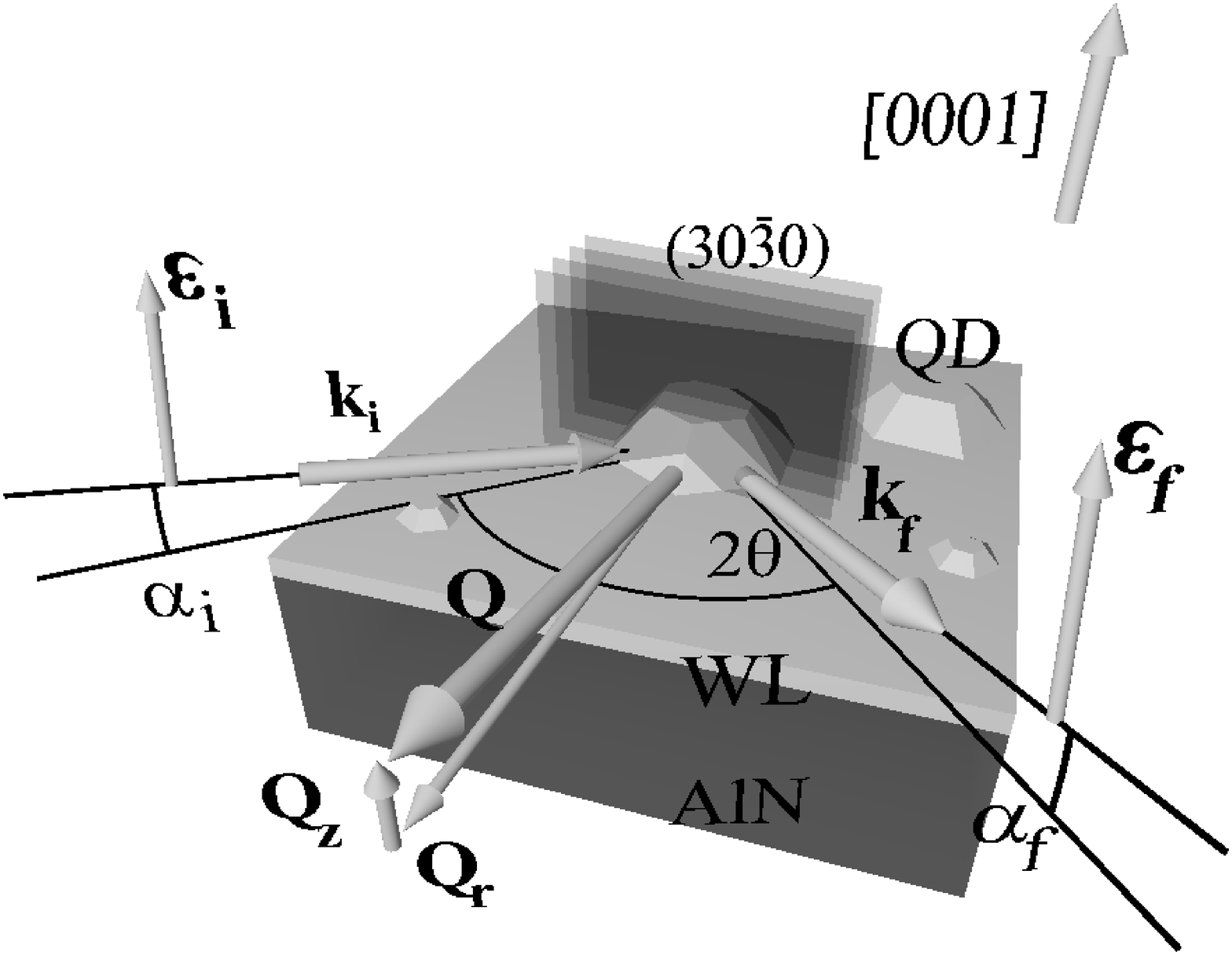}\end{center}
\caption{\label{fig:gid}}
\end{figure}

Grazing Incidence Diffraction Anomalous Fine Structure
(GIDAFS) at the Ga K-edge (10.367 keV) was performed at the French
Collaborative Research Group beamline BM2 at the European Synchrotron
Radiation Facility (ESRF) by using the 8-circle diffractometer equipment.
We measured the diffuse scattering intensity, in grazing incidence
and exit, close to the in-plane $(30\bar{3}0)$ Bragg reflection of
the AlN substrate (radial scans), at energies close to the Ga K-edge.
Figure \ref{fig:gid} sketches the experimental set-up. The samples
were mounted in the vertical plane, i.e. the polarization vector $hat{\varepsilon_{i}}$
of the incident photon beam was perpendicular to the sample surface
(0001). The incidence angle was $\alpha_{i}=0.17^\circ,$ lower than the
bulk AlN critical angle $\alpha_{c}=0.21^\circ$ (at $10.32\: keV$) for
which the total reflection regime takes place. Such conditions were used to enhance the weak
contribution of the encapsulated dots layer with respect to that of
the substrate. The diffraction
geometry was chosen in such a way to keep the scattering vector in
the vertical plane. The diffraction point-detector was a scintillator and
the slits were opened so as to measure the integrated intensity over
the grazing exit angle $\alpha_{f}$. A photodiode measuring the fluorescence
yield of an in-vacuum 4 $\mu m$ Ti-foil was used to monitor the incoming x-ray
beam. Two kinds
of scans were performed : a) \textit{h}-scans (radial)
in the range 2.9-3.05 for 12 energies, from 10.272 to 10472 keV, i.e.
close to the Ga K-edge, and b) energy
scans at fixed scattering vector (\textbf{Q})
corresponding to the maximum of the QDs contribution to the diffuse
scattering (i.e. at the maximum of the partial structure factor $F_{A=Ga}$
profile extracted from the multiwavelength \textit{h}-scans,
see the following section and ref. \cite{Letoublon04}). The energy
scans were recorded in a large energy interval, typically 1 keV ,
with an energy step from 1 to 2 eV, to allow the quantitative analysis
of both the edge and the extended oscillations.

\section{MAD analysis}

\label{sec:Multiwavelength-Anomalous-Diffraction}

\begin{figure}
\begin{center}\includegraphics[width=10cm,keepaspectratio]{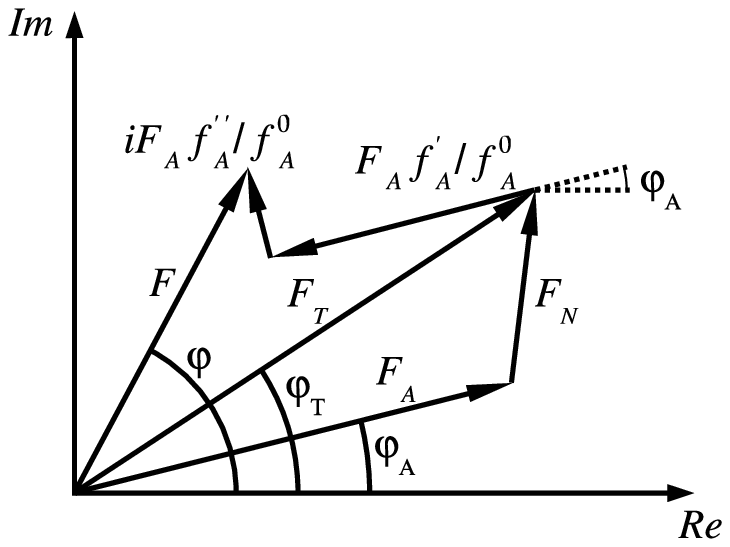}\end{center}
\caption{\label{fig:Structure_factor}}
\end{figure}

The solid lines in fig. \ref{fig:MAD_extraction}(a,c,e)
show square root intensities along \textit{$\left[10\bar{1}0\right]$}
direction (\textit{h}-scan)
close to the $\left(30\bar{3}0\right)$ reflection
as a function of the AlN coverage. These \textit{h}-scans
are related to both the in-plane strain state and size. With no AlN
coverage, one observes a diffuse-scattering peak ascribed to QDs slightly
strained by the AlN buffer and substrate. As the AlN coverage increases
(from (a) : free standing QDs to (e) : 10 MLs AlN coverage), this
peak is progressively shifted towards higher \textit{h}
values and gets mixed to the AlN buffer peak. Further
analysis was made possible by distinguishing the GaN and AlN contributions,
using MAD measurements \cite{Letoublon04,Hodeau01,Hendrickson91}.
Figures \ref{fig:MAD_extraction}(b,d,f) show some of the square root
diffracted intensities measured for increasing AlN coverage, across the Ga K-edge, taking advantage of the Ga anomalous
effect to localize the Ga contribution along $\left[10\bar{1}0\right]$.
The Ga scattering factor can be written as $f_{Ga}=f_{Ga}^{0}+f_{Ga}^{\prime}+if_{Ga}^{\prime\prime}$,
where $f_{Ga}^{\prime}$ and $f_{Ga}^{\prime\prime}$ are the Ga real
and imaginary anomalous (resonant) scattering corrections, $f_{Ga}^{0}$
is the Ga Thomson scattering factor. From MAD measurements, the Ga
partial structure factor $F_{Ga}$ of phase $\varphi_{Ga}$, that
includes the Thomson scattering of all anomalous atoms (Ga), can be
retrieved. The retrieval shall be run in the framework
of the Distorted Wave Born Approximation, taking into account scattering paths
involving the reflection from the layer supporting the dots
\cite{Pietsch04,Schmidbauer05}. We recorded the diffracted intensity integrated
over the exit angle $\alpha_f$, between 0 and $2\alpha_c$, and therefore
collected all the scattering paths. Discarding the energy dependence of the reflection coefficients
at the Ga K-edge, as a consequence of the small Ga amount (6 equivalent
MLs), the recorded intensity corrected for fluorescence, $I_{exp}$, is proportional to the
total square structure factor $\left\Vert F\right\Vert ^{2}$ :

\begin{equation}
I_{exp}\left(E\right)\propto\left\Vert F\right\Vert ^{2}\propto\left\Vert F_{T}\right\Vert ^{2}\times\left\lbrace \left[\mathrm{cos}\left(\varphi_{T}-\varphi_{Ga}\right)+\beta f'_{Ga}\right]^{2}+\left[\mathrm{sin}\left(\varphi_{T}-\varphi_{Ga}\right)+\beta f''_{Ga}\right]^{2}\right\rbrace \label{eq:Iexp}\end{equation}

where $\beta=\left\Vert F_{Ga}\right\Vert /\left(f_{Ga}^{0}\left\Vert F_{T}\right\Vert \right)$.
Figure \ref{fig:Structure_factor} shows the total and partial structure
factors relations in the complex plane. The partial structure factor
$F_{T}$ of phase $\varphi_{T}$ that includes the overall contribution
of non anomalous atoms and the Thomson scattering of all anomalous
atoms, $F_{Ga}$, as well as $\varphi_{T}-\varphi_{Ga}$, can be extracted
for all \textit{h} values, without
any structural model by fitting eq. \ref{eq:Iexp} to the experimental
data with the \texttt{NanoMAD} algorithm
\cite{FavreNicolin05}. 

As shown in fig. \ref{fig:MAD_extraction}(a,c,e),
$F_{Ga}$ and $F_{T}$ were extracted. The $h=h_{Ga}$
position of the diffuse $F_{Ga}$ peak maximum is
inversely proportional to the in-plane average lattice parameter $a_{GaN}$,
since the distance between GaN $\left(30\bar{3}0\right)$ planes is
$d_{30\bar{3}0}=\sqrt{3}/2 \times a_{GaN}/3=\sqrt{3}/2 \times a_{AlN}/h_{Ga}$
with $a_{AlN}\simeq3.112 \AA$, for the AlN substrate peak used as
a reference. Figure \ref{fig:a_GaN} shows the evolution of the in-plane
lattice parameter as a function of the AlN cap thickness. The uncapped
QDs are partially in-plane relaxed, with an average strain relative
to bulk GaN, $\varepsilon_{xx,GIXRD}=\left(a_{GaN,GIXRD}-a_{GaN,bulk}\right)/a_{GaN,bulk}\sim-1\%$,
with $a_{GaN,bulk}=3.189\AA$. The QDs are then progressively in-plane
compressed by the AlN capping, but remain slightly relaxed, $\varepsilon_{xx,GIXRD}\sim-1.6\%$,
compared to pseudomorphic GaN ($\varepsilon_{xx}\sim-2.4\%$).

\section{EDAFS analysis}

\label{sec:Extended-DAFS-analysis}

Out-of-plane information can be achieved by quantitative
analysis of the Grazing incidence Diffraction Anomalous Fine Structure
(GIDAFS) oscillations in the extended region above the edge (EDAFS).
Figure \ref{fig:1967_chi and FT}(a) shows for free
standing QDs the oscillatory contribution $(\chi_{DAFS})$ to the
DAFS spectrum, extracted and normalized to the smooth atomic background
($I_{0}$) :\[
\chi_{DAFS}=\frac{I_{exp}-I_{0}}{I_{0}}\]

$\chi_{DAFS}$ can be written as $\chi_{DAFS}=\frac{1}{S_{D}}\chi_{Q}$
where $S_{D}$ is a normalization factor that depends on crystallography
and is calculated from the parameters $\bigtriangleup\varphi=\varphi_{T}-\varphi_{A}$
and $\beta$ (see eq. \ref{eq:Iexp}) , $\chi_{Q}$ is in the first order
approximation of the diffracted anomalous fine structure \cite{Proietti99}
:

\begin{equation}
\chi_{\mathbf{Q}}(k)=cos(\varphi-\varphi_{A})\sum_{j=1}^{N_{A}}w_{j}^{\prime}\chi_{j}^{\prime}+sin(\varphi-\varphi_{A})\sum_{j=1}^{N_{A}}w_{j}^{\prime\prime}\chi_{j}^{\prime\prime}\label{eq:chi_Q}\end{equation}

where the \textit{j} label
runs over the different anomalous sites A (the Ga sites, A=Ga), $w_{j}^{\prime}=\frac{\left\Vert F_{j}\right\Vert cos(\varphi-\varphi_{j})}{\left\Vert F_{A}\right\Vert cos(\varphi-\varphi_{A})}$
and $w_{j}^{\prime\prime}=\frac{\left\Vert F_{j}\right\Vert sin(\varphi-\varphi_{j})}{\left\Vert F_{A}\right\Vert sin(\varphi-\varphi_{A})}$
are crystallographic weights. The term $\chi_{j}^{\prime}$ ($\chi_{j}^{\prime\prime}$)
in eq. \ref{eq:chi_Q} is the oscillatory part of the resonant atomic
scattering factor $f_{j}^{\prime}$ ($f_{j}^{\prime\prime}$), it
is related to the local atomic environment of the resonant atom. $\chi_{j}^{\prime\prime}$
is formally identical to the Extented X-ray Absorption Fine Structure
(EXAFS) oscillations of atom $j$. In the present case, i.e. one statistical
equivalent site, $\chi_{Q}$ can be rewritten as a function of the
virtual photoelectron wave vector modulus \textit{k} in a form that is similar
to the well known EXAFS formula \cite{Proietti99} :

\begin{equation}
\chi_{\mathbf{Q}}(k)=\sum_{\gamma}A_{\gamma}(k)sin\left[2k\left\langle R\right\rangle _{\gamma}+\varphi_{\gamma}(k)+2\delta_{c}(k)+\varphi-\varphi_{A}-\frac{\pi}{2}\right]\label{eq:chi_Q_XAFS_equivalent}\end{equation}
 where $\gamma$ runs over all the possible virtual photoelectron
scattering paths, $\left\langle R\right\rangle _{\gamma}$ is the
effective length of path $\gamma$, $\varphi_{\gamma}(k)+2\delta_{c}(k)$
is the net scattering photoelectron phase shift.

The analysis can be performed according to the standard
criteria and available codes for EXAFS, provided that crystallographic
phases and amplitude correction factors are taken into account (for
more details see references \cite{Proietti99,Renevier03}). The EDAFS
analysis has been carried out by using the \texttt{\texttt{FEFF8}}
code \cite{Ankudinov98} to generate theoretical phases
and amplitudes, taking into account beam polarization, for a $6\AA$
radius GaN cluster. In order to address for the possible presence
of Al atoms in the QDs or at the substrate and capping interface,
Ga-Al and Ga-N-Al scattering paths were considered by calculating
an AlN cluster with the Ga central atom as absorber. The \texttt{Artemis}
interface to the \texttt{IFEFFIT}
package \cite{Newville95} was used to fit theoretical
computations to the experimental data.

\begin{figure}
\begin{center}\includegraphics[width=12cm,keepaspectratio]{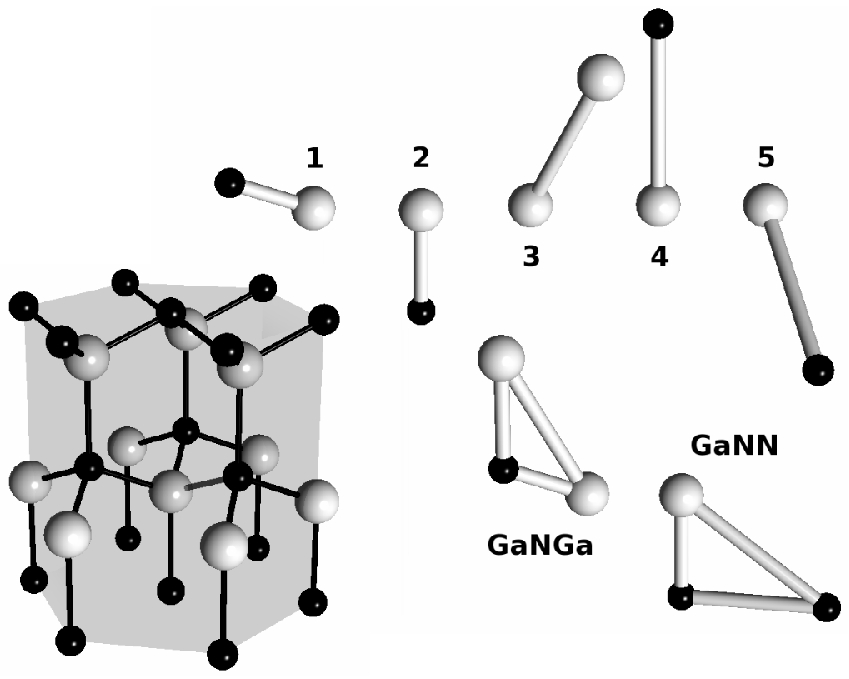}\end{center}
\caption{\label{fig:paths}}
\end{figure}

The EDAFS spectra were Fourier Transformed in the
k-range 3-10 $\AA^{-1}$, and the fit was performed in R-space
(real space), using 4 next neighboring shells (I-IV). As an example,
we show the best fit curves for free standing QDs in fig. \ref{fig:1967_chi and FT}(a)
and \ref{fig:1967_chi and FT}(b), compared to the experimental raw
data. Six Single Scattering paths (SS) and four 4 Multiple Scattering
(MS) paths were found to be relevant in this range (see fig. \ref{fig:paths})

\begin{enumerate}
\item $(Ga-N)_{//}$, in-plane, I shell path, corresponding
to the 3 Ga-N bonds of the tetrahedron that are nearly in-plane,
\item $(Ga-N)_{\perp}$, out-of-plane, I shell path, corresponding
to the fourth Ga-N bond of the tetrahedron, lying along \textit{c}-axis,
\item $Ga-Ga$, II shell, out-of-plane path, corresponding
to 6 Ga atoms at a distance that is a combination of $a$ and $c$,
$\left\{ \frac{1}{3}a_{GIXRD}^{2}+\frac{1}{4}c^{2}\right\} ^{1/2}$,
where $a_{GIXRD}$ is the in-plane lattice parameter obtained with
the Grazing Incidence (and exit) X-ray Diffraction experiment (see
\ref{sec:Multiwavelength-Anomalous-Diffraction}),
\item$Ga-N$, III shell path, corresponding to one N
atom along $c$ direction,
\item$Ga-N$, IV shell path, corresponding to 6 N nearly
in-plane atoms,
\item MS paths consisting of triangular paths Ga-N-N and
Ga-N-Ga.
\end{enumerate}
The "in-plane" statement refers to
the surface or growth plane, and all the scattering paths, except
the first one, were expressed in terms of \textit{a}
and \textit{c} cell
parameters, as requested by the hexagonal cell symmetry.

We performed the fit by fixing the \textit{a}
parameter (in-plane) to the values found by diffraction,
$a_{GaN,GIXRD}$, letting the \textit{c} parameter
vary according to the hexagonal symmetry. The \textit{Ga-N} first shell distances were let free to vary independently
of \textit{a} and \textit{c} since, as it is well known, the Vegard's law is
far from being valid for semiconductor alloys, in which the bond-bending
mechanism is dominant compared to bond-stretching \cite{Romanato98}.
The presence of Al is taken into account by adding the correspondent
scattering paths in which Al substitutes for Ga as Next Nearest Neighbour
(NNN) and mutiplying the amplitude by a factor $x_{Al}$ for Al and ($1-x_{Al}$)
for Ga. The best fit parameters are shown, for the whole set of samples
in table \ref{tab:table}, where we also report, as a reference,
the bulk and pseudomorphic values for GaN \cite{Schulz97}. The Ga-Al
distance was also let free to vary and the values found were close
to the Al-Al NNN distance. Since the Al content is found to be zero
within the statistical errors, the Al-Al NNN distance was not reported
in the table. Starting from the fit results, the in-plane and out-of-plane
strain were calculated, with respect to relaxed (bulk) GaN, as $\varepsilon_{xx,GIXRD}=\left(a_{GaN,GIXRD}-a_{GaN,bulk}\right)/a_{GaN,bulk}$
and $\varepsilon_{zz,GIDAFS}=\left(c_{GaN,GIDAFS}-c_{GaN,bulk}\right)/c_{GaN,bulk}$.
$\varepsilon_{xx,GIXRD}$ \textit{vs} $\varepsilon_{zz,GIXRD}$
for the different samples studied are sketched in fig. \ref{fig:e_xx and e_zz}.
These values are compared to the biaxial elastic behaviour for pseudomorphic
GaN on AlN (straight line), which corresponds to $\varepsilon_{xx}=-2\varepsilon_{zz}c_{13}/c_{33}$,
with the elastic coefficients $c_{13}$ and $c_{33}$ values from
reference \cite{Andreev00}.

We observe the following general findings :

\begin{enumerate}
\item the Ga-N first shell in-plane and out-of-plane
distances show to be very close to each other, within the fit errors
(0.01 $\AA$), in agreement with previous studies \cite{Acapito02},
\item as shown in table \ref{tab:table}, the Al
content remains very small, showing that no intermixing takes place
in the QDs as expected for the Al/Ga species \cite{Arlery99},
\item the $c_{GaN,GIDAFS}$ values range from 5.22 to 5.25
$\AA$, that is quite large compared to the values foreseen
by the elastic regime of a pseudomorphic GaN layer, as apparent in
fig. \ref{fig:e_xx and e_zz} where the experimental points fall
above the elasticity curve.
\end{enumerate}
As expected, the uncapped QDs do not follow a biaxial strain behaviour, in
reason of the presence of free surface. However, capping by a thin layer (2-5 MLs) of AlN should favor the evolution towards the biaxial case for at least two reasons : a) first
of all, capping results in a size decrease of the dots, i.e. an aspect ratio reduction associated with a relative increase of the
biaxial component of the strain \cite{Gogneau04} ; b) furthermore we speculate that the possible wetting of the QDs by AlN, which will
be discussed in section \ref{Edge analysis}, strongly tends to reduce the relaxation through the free surface, which also results in a
relative increase of the biaxial component of the strain. Besides these considerations, it is worth noting that the plastic relaxation
process of AlN deposited on GaN which is characterized by a very small critical thickness \cite{Bourret01}, is still unclear and may
also determine to some extent the strain state of the AlN/GaN QDs interacting system.

\section{Edge analysis}

\label{Edge analysis}

The diffraction anomalous spectra, close to the
Ga K-edge, can give the Al and Ga relative composition inside the
GaN/AlN in-plane iso-strain region selected with grazing incidence
and exit diffraction, this region includes the GaN QDs and the AlN
on top. Previous studies clearly indicates that no atomic intermixing
occurs at the GaN/AlN interfaces, neither with GaN/AlN QWs nor GaN/AlN
QDs \cite{Arlery99,Gogneau04}. This is confirmed by the grazing incidence
EDAFS results reported in section \ref{sec:Extended-DAFS-analysis}
that clearly show no significant Al/Ga mixing.

\begin{figure}
\begin{center}\includegraphics[width=12cm,keepaspectratio]{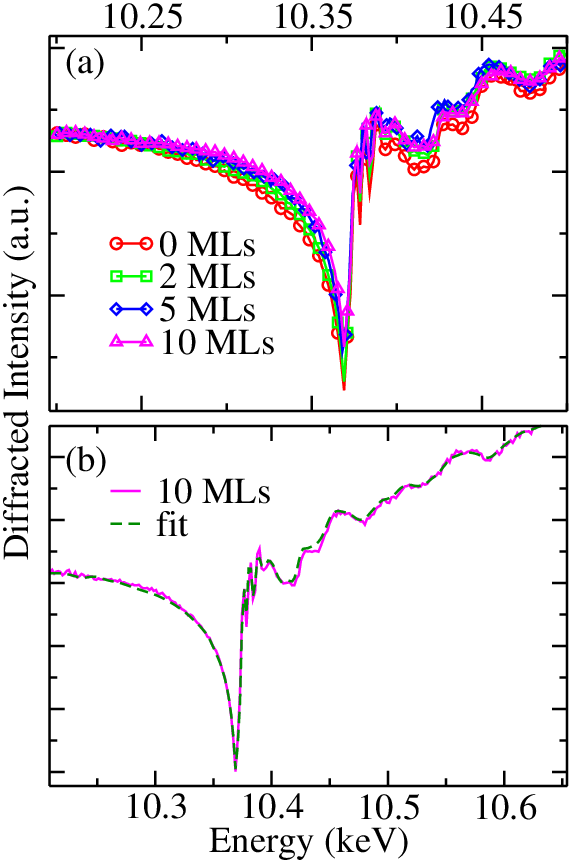}\end{center}
\caption{\label{fig:DANES_comp and dafs_1957}}
\end{figure}

\begin{figure}
\begin{center}\includegraphics[width=12cm,keepaspectratio]{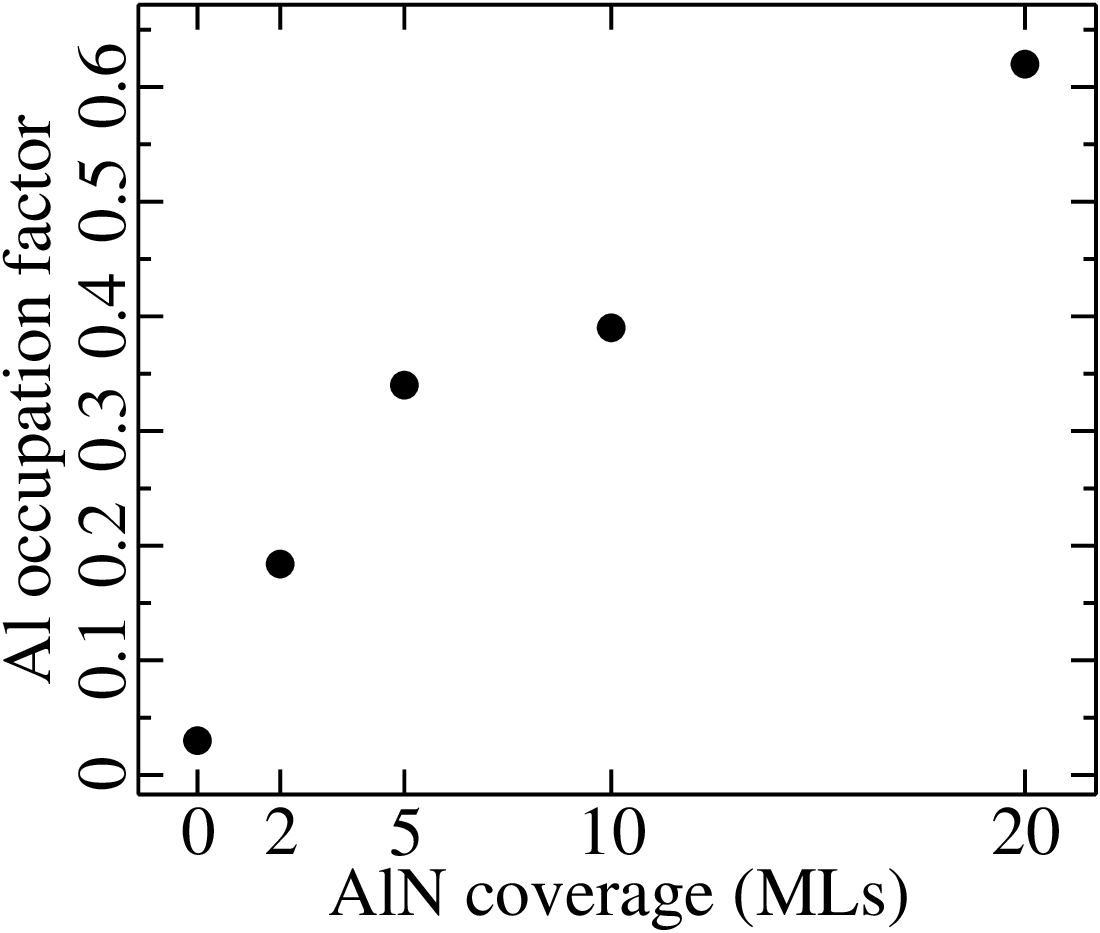}\end{center}
\caption{\label{fig:AlN_x}}
\end{figure}

On the other side, the analysis of the DAFS edge shape can give information
about the capping mechanism of the QDs. Indeed, as a first approximation, the
diffracted intensity is proportional to the square modulus of the in-plane
iso-strain region structure factor. We calculate this structure factor for an
$Al_xGa_{1-x}N$ wurtzite structure to take into account the Al atoms belonging to the
same iso-strain region as the Ga atoms at the QDs top. The Al concentration
obtained by refining the $x$ value gives the Al atoms fraction seen by
diffraction, contributing at the chosen Q-value, and determining the edge line
shape. Taking into account that EDAFS analysis shows that no intermixing takes
place, we cane state that we are probing the AlN capping. Figure \ref{fig:DANES_comp and dafs_1957}a) shows the GIDAFS spectra
for the 0, 2, 5 and 10 AlN MLs cap thicknesses, measured at the maximum
of the partial structure factor $F_{A=Ga}$ (see section \ref{sec:Multiwavelength-Anomalous-Diffraction}).
The data were normalized so that the intensity at 10.2 keV is the
same for all spectra. Equation \ref{eq:Iexp} was fitted to each GIDAFS
spectrum, using the anomalous scattering factors $f_{Ga}^{\prime}$
and $f_{Ga}^{\prime\prime}$ of a GaN layer. A scale factor, the detector
efficiency as a function of the energy and the Al occupation factor
($x$) inside the in-plane iso-strain
region were refined. As an example, fig. \ref{fig:DANES_comp and dafs_1957}(b)
shows the best fit for the 10 MLs sample obtained with $x=0.39\pm0.01$.
It should be noted that the occupation factor is determined by the
ratio $\beta=\frac{\left\Vert F_{A}\right\Vert }{f_{Ga}^{0}\left\Vert F_{T}\right\Vert }$,
i.e. the curvature and depth of the cusp before and at the edge.
The fit quality is the same for all samples. Figure \ref{fig:AlN_x}
shows the Al concentration ($x$)
as a function of the AlN cap thickness. Up to 5 MLs, the AlN contribution,
that is near to zero ($x=0.3\pm0.01$)
for free standing QDs, increases linearly, stabilizes above
5MLs, and at 10 MLs the contribution is almost
the same as for the 5 MLs sample. Provided that AlN on top of the
QDs is pseudomorphic to GaN for low coverages \cite{Coraux05a}, the evolution of
the Al concentration ($x$) in the iso-strain region up to 5 MLs indicates a
linear increase of the amount of AlN on top of the GaN QDs. This could result
from a uniform growth of AlN on the whole surface, i.e. the QDs and the thin pseudomorphic
wetting layer \cite{Gogneau04}. The further evolution of the Al concentration
above 5 MLs points out a change in the AlN growth process, leading to AlN with
an in-plane strain state different from that in the QDs. This change may correspond
either to plastic relaxation in AlN, or to selective AlN growth in
between of the QDs where the stress differs from that on top of the QDs, or to another
still unknown process.

\section{Conclusion}

We have presented new results on the structural properties
of GaN QDs by combining different aspects of X-ray diffraction : quasi
surface sensitivity due to grazing incidence, quantitative analysis
of anomalous effect according to MAD principles, lineshape fit of
DAFS and EDAFS oscillations fit.

All these aspects are strongly complementary. We
determine in-plane and out-of-plane lattice parameters and investigate
the effect of capping layer by monitoring its effect on the QDs strain.
In addition, the Al fraction seen by the anomalous diffraction as
a function of the capping layer thickness (obtained by the GIDAFS
lineshape analysis at the Ga K-edge), indicates a wetting of the
QDs, followed by a noticeable change in the capping process which may be related
either to plastic relaxation in AlN, or to spatially selective AlN growth, or to
a still unknown process. Let us point out that the Al
fraction obtained in this way does not represent the Al content inside
the dots, but the AlN contribution to the diffuse scattering at the
same \textbf{Q} value as the
GaN QDs contribution, i.e. AlN mostly located on top of the QDs. The
Al content of the dots can be found by analysis of the EDAFS oscillations
which provide the microscopic local environment of the Ga resonant
atom. Our analysis shows that no Ga/Al intermixing takes place, as
expected for these two group III-N elements.
We recovered the in-plane and perpendicular strains $\varepsilon_{xx}$
and $\varepsilon_{zz}$ in the dots and compare them to the biaxial
elastic strain of a pseudomorphic layer. We find a tendency to an
over-strained regime that suggests a more complex mechanism of strain
accomodation which deserves further investigations.

We are very grateful to N. Boudet, S. Arnaud, B. Caillot and J.F.
B$\acute{e}$rar for help to set-up the GIDAFS experiment at beamline BM2 at
the ESRF. We are very grateful to P. Wolfers who gave us the DPU code
used for data analysis. The authors would like to thank C. Priester
for her critical reading of the manuscript. 
MGP acknowledges the support of the Spanish Ministry of Education and Science in the frame of "Programa de estancias de
profesores espa$\tilde{n}$oles en centros de investigaci$\acute{o}$n extranjeros" (project n. PR2005-0231).
\bibliographystyle{unsrt}
\bibliography{GIDAFS_prb}

\newpage

\section*{Figure Captions}

Fig. \ref{fig:AFM_2D} : $1\mu m^{2}$ AFM image of uncapped GaN QDs grown
in the modified SK mode.

Fig. \ref{fig:gid} : Grazing incidence geometry for a in-plane reflection
$\left(30\bar{3}0\right)$. See text for details.

Fig. \ref{fig:MAD_extraction} : (a,c,e) : $\sqrt{I_{exp}}$ measured
at 10.317 keV (50 eV below Ga K-edge), $\left\Vert F_{Ga}\right\Vert $
and $\left\Vert F_{T}\right\Vert $ extracted for a 0 MLs (a), a 5
MLs (c), and a 10 MLs (e) AlN coverage. (b,d,f) : Experimental square
root intensities $\sqrt{I_{exp}}$ measured below (-100 and -50 eV),
at (edge), and above (+ 5eV) the Ga K-edge for a 0 MLs (b), a 5 MLs
(d), and a 10 MLs (f) AlN coverage. 

Fig. \ref{fig:Structure_factor}: Schematic representation in the
complex plane of the structure factor $F$ as a function of $F_{T}$,
$F_{A=Ga}$ and $\varphi_{T}-\varphi_{A=Ga}$ (see text). $F_{N}$
represents the partial structure factor of non resonant atoms (Al,
N).

Fig. \ref{fig:a_GaN} : In-plane lattice parameter $a_{GaN,GIXRD}$
and strain (relative to bulk GaN) in GaN deduced from the position
of the $F_{Ga}$ maximum. Bulk GaN gives $\varepsilon_{xx,GIXRD}=0\%$
with $a_{GaN,bulk}=3.189\AA$ while bulk AlN gives $\varepsilon_{xx,GIXRD}=-2.4\%$
with $a_{AlN}=3.112\AA$.

Fig. \ref{fig:1967_chi and FT} :  (a) Experimental EDAFS for the free
standing QDs sample, compared with the best fit result, (b)
R-space experimental curve for free standing QDs compared with best fit.

Fig. \ref{fig:paths} :  Scheme of GaN wurzite structure, the
most relevant virtual phoelectron scattering paths used for the EDAFS
simulation are represented : (1) in-plane I shell $(Ga-N)_{//}$,
(2) out-of-plane I shell $(Ga-N)_{\perp}$, (3) out-of-plane
II shell $(Ga-Ga)_{\perp}$, (4) III shell \textit{Ga-N} along \textit{c},
(5) nearly in-plane IV shell \textit{Ga-N}, MS Ga-N-N and Ga-N-Ga. Ga atoms are represented by white spheres, N by black ones.

Fig. \ref{fig:e_xx and e_zz} :  GaN QDs strain $\varepsilon_{xx} =\left(a_{GaN,GIXRD}-a_{GaN,bulk}\right)/a_{GaN,bulk}$
\textit{versus} $\varepsilon_{zz}=\left(c_{GaN,GIDAFS}-c_{GaN,bulk}\right)/c_{GaN,bulk}$ values
for all the samples studied compared with elastic biaxial strain of
a pseudomorphic GaN thin film.

Fig. \ref{fig:DANES_comp and dafs_1957} : (a) GIDAFS spectra for
0, 2, 5 and 10 MLs AlN capping, measured at maximum of $F_{A}$; (b)
crystallographic best fit for the 10 ML AlN sample, open circle :
experiment, solid line : simulation performed with experimental $f_{Ga}^{\prime}$
and $f_{Ga}^{\prime\prime}$ of a GaN thin film.

Fig. \ref{fig:AlN_x} : Al atoms occupation factor, $x$, of the $Al_{x}Ga_{1-x}N$
iso-strain region selected by diffraction (maximum of $F_{Ga}$),
as a function of AlN coverage.

\section*{Table caption}

Table \ref{tab:table} : EDAFS best fit values for interatomic
distances (R), Debye-Waller factors ($\sigma$) and Al concentration
($x_{Al}$) obtained by \texttt{IFEFIT} mimimization using theoretical fitting
standards provided by \texttt{FEFF8} code. The amplitude and phase
correction factors have been obtained by crystallographic analysis
of the DAFS lineshape. The $a_{GaN,GIXRD}$ value is kept fixed to
the value determined by grazing incidence and exit diffraction (diff.).

\newpage

\begin{table}
\caption{\label{tab:table}}
\begin{center}\begin{tabular}{|c|c|c|c|c|c|c|}
\hline 
&
 Bulk&
 GaN/AlN&
 0 MLs&
 2 MLs&
 5 MLs&
 10 MLs\tabularnewline
\hline
R1(Ga-N) ($\AA$)&
 -&
 -&
 1.93&
 1.94&
 1.94&
 194\tabularnewline
\hline
$\sigma_{1}^{2}(\AA^2)$&
 -&
 -&
 $2\times10^{-3}$&
 $4\times10^{-3}$&
 $4\times10^{-3}$&
 $1\times10^{-3}$\tabularnewline
\hline
R2 (Ga-Ga)$_{//} = a_{GaN}$ ($\AA$)&
 3.188&
 3.11&
 3.156 (diff.)&
 3.147 (diff.)&
 3.149 (diff.)&
 3.14 (diff.)\tabularnewline
\hline
$\sigma_{1}^{2}(\AA)^{2}$&
 -&
 -&
 $6\times10^{-3}$&
 $8\times10^{-3}$&
 $4\times10^{-3}$&
 $7\times10^{-3}$\tabularnewline
\hline
R2 (Ga-Ga) ($\AA$)&
 3.18&
 -&
 3.19&
 3.18&
 3.18&
 3.19\tabularnewline
\hline
$c_{GaN}$ $(\AA$)&
 5.186&
 5.26&
 5.25$\pm0.02$&
 5.23$\pm0.03$&
 5.22$\pm0.02$&
 5.25$\pm0.04$\tabularnewline
\hline
$c_{GaN}/a_{GaN}$&
 1.626&
 1.69&
 1.66&
 1.66&
 1.66&
 1.67\tabularnewline
\hline
$x_{Al}$&
 -&
 -&
 0.1$\pm0.1$&
 0.0$\pm0.1$&
 0.1$\pm0.1$&
 0.05$\pm0.1$ \tabularnewline
\hline

\end{tabular}\end{center}
\end{table}
\newpage

\newpage

\newpage

\begin{figure}
\begin{center}\includegraphics[width=15cm,keepaspectratio]{extraction_bis.eps}\end{center}
\caption{\label{fig:MAD_extraction}}
\end{figure}
\newpage

\newpage

\begin{figure}
\begin{center}\includegraphics[width=12cm,keepaspectratio]{aGaN.eps}\end{center}
\caption{\label{fig:a_GaN}}
\end{figure}
\newpage

\begin{figure}
\begin{center}\includegraphics[width=15cm,keepaspectratio]{fit_s1967_chi_ft.eps}\end{center}
\caption{\label{fig:1967_chi and FT}}
\end{figure}
\newpage

\newpage

\begin{figure}
\begin{center}\includegraphics[width=12cm,keepaspectratio]{exx_ezz_01_07_MEIS.eps}\end{center}
\caption{\label{fig:e_xx and e_zz}}
\end{figure}
\newpage

\newpage

\newpage

\end{document}